\documentclass[twocolumn,showpacs,preprintnumbers,amsmath,amssymb]{revtex4}
\usepackage{graphicx}
\usepackage{dcolumn}
\usepackage{bm}

\begin{document}

\preprint{APS/123-QED}
\title{Cavity-enhanced superradiant Rayleigh scattering with ultra-cold and Bose-Einstein condensed atoms}

\author{Sebastian Slama}
\author{Gordon Krenz}
\author{Simone Bux}
\author{Claus Zimmermann}
\author{Philippe W. Courteille}
\affiliation{Physikalisches Institut, Eberhard-Karls-Universit\"at T\"ubingen,\\
Auf der Morgenstelle 14, D-72076 T\"ubingen, Germany}

\date{\today}

\begin{abstract}
We report on the observation of collective atomic recoil lasing and
superradiant Rayleigh scattering with ultracold and Bose-Einstein
condensed atoms in an optical ring cavity. Both phenomena are based
on instabilities evoked by the collective interaction of light with
cold atomic gases. This publication clarifies the link between the
two effects. The observation of superradiant behavior with thermal
clouds as hot as several tens of $\mu\textrm{K}$ proves that the
phenomena are driven by the cooperative dynamics of the atoms, which
is strongly enhanced by the presence of the ring cavity.
\end{abstract}


\maketitle

\section{Introduction}

The interaction of light with atomic gases takes place in most cases as a local process: Light shone into an atomic cloud is scattered by individual atoms. In principle, every atom having scattered a photon can be detected through the momentum imparted to it by photonic recoil, and in general, the scattering process is ignored by all other atoms. This holds even for Bose-Einstein condensates (BEC), which are pure quantum states consistent of an ensemble of delocalized atoms. There are however prominent exceptions: Dicke superradiance \cite{Dicke54} is a well-known synchronization phenomenon in spontaneous emission. It is observed, for instance, as a collective deexcitation of an ensemble of inverted atoms with an accelerated rate, which scales with the square of the number of inverted atoms \cite{Skribanowitz73}. Another example is the collective absorption of photonic recoil by an ensemble of atoms tight together by strong forces known as M\"o\ss bauer effect \cite{Moessbauer58,Steane97}.

Collective effects in light scattering arise when the scatterers are mutually coupled by interactions or display long-range order. Often the collective coupling involves mechanical forces, for example photonic recoil or the electrostrictive force arising from dipole-dipole interactions. In both cases, the interatomic force originates from a radiative interaction, or using fully quantized terms, the transfer of phonons is mediated by an exchange of photons. Compared to short-ranged binary collisions radiation-based interaction extends much further in space. Under some circumstances it can be completely delocalized. In some cases, collective coupling can trigger instabilities. Well-known examples for instabilities in the field of nonlinear optics are stimulated Raman scattering, stimulated Brillouin scattering or the collective atomic recoil laser (CARL) \cite{Bonifacio94,Bonifacio95,Kruse03b,Cube04,Slama07}.

Collective instabilities have recently been observed in clouds of cold and ultra-cold atoms driven by light \cite{Inouye99,Kozuma99,Fallani05,Kruse03b,Elsaesser03b,Black03,Labeyrie06}. In the present paper, we focus on two types of experiments,
dealing with the superradiant Rayleigh scattering (SRyS) phenomenon on one hand \cite{Inouye99,Kozuma99,Fallani05} and the collective atomic recoil laser \cite{Kruse03b} on the other.

CARL is observed, when a strong pump field is shone onto an atomic gas. This leads to the exponential growth of an unpumped probe light field and to the formation of an atomic density grating \cite{Bonifacio94,Bonifacio95}. If pump and probe light field are counter-propagating modes of a high-finesse ring cavity, the interaction time of the light fields with the atoms can be enhanced by several orders of magnitude, which supports the amplification. Consequently, all CARL experiments carried out up to date employed ring cavities \cite{Kruse03b,Cube04,Slama07}. Therefore, in this paper we will use the term CARL in the tight sense of a cavity-assisted collective instability, although the CARL has originally been postulated without cavity \cite{Bonifacio94}.

SRyS has first been observed in Bose-Einstein condensed atomic
clouds. A short laser pulse shone onto the cloud is scattered from
atoms of the BEC, which then by photonic recoil form motional
sidemodes. Matter-wave interference between the recoiling atoms and
the BEC at rest leads to the formation of an atomic density grating
thereby exponentially enhancing the scattering. SRyS was originally
attributed to four-wave mixing between optical and matter waves,
bosonically stimulated by the macroscopic occupation of the final
momentum state. Already in the pioneering work \cite{Inouye99} it
was recognized that SRyS does not require quantum degeneracy and
would in principle also work in a thermal cloud. Nevertheless the
terminology of bosonic stimulation and the fact that SRyS could at
first not be observed with thermal clouds led to some obscurity and
discussions about the role of quantum statistical effects. Theoretic
work \cite{Moore01,Ketterle01} showed that the gain mechanism is
independent of the quantum statistics and should in principle also
be observable with fermionic and thermal atoms. The experimental
prove was given by the observation of CARL \cite{Kruse03b} and SRyS
with thermal gases \cite{Yoshikawa05}. The important feature is not
the quantum state of the atoms but the cooperative behavior.

CARL has a close analogy with SRyS, since they both share the same gain mechanism \cite{Piovella01}. However in contrast to SRyS, CARL activity has been observed with thermal atoms as hot as a few $100~\mu$K \cite{Kruse03b}. This fact raises the question, what distinguishes both collective effects. In both experiments there must be a coherent mechanism correlating the individual scattering events. Coherence can be transferred between scattering events either via de Broglie waves interference or optical interference.

SRyS is difficult to observe with thermal atomic ensembles, because
the coherence is stored in the momentum states of the atoms. Thermal
motion of atoms therefore Doppler-limits the coherence time of the
system \cite{Yoshikawa05}. CARL is much less sensitive to the
thermal motion of the atoms, because the coherence is stored in the
light field of the cavity. The density-of-states in the cavity
restricts the frequency of the scattered light to values close to
one of its eigenfrequencies. In the case of a so-called good-cavity
this is equivalent to the fact that the atomic momentum states which
can be populated by photonic momentum transfer are limited to a few
low-lying states. This effect counteracts momentum diffusion which
can occur due to a thermal motion of the atoms, but is also
intrinsically connected with the collective gain process itself.

\bigskip

We organized this paper as follows: In section~\ref{SecTheory} we expose the problem of motion-induced collective effects in light scattering. In particular, we will discuss the intricate relationship between CARL and SRyS, pointing out the common features and the differences. We will then briefly introduce the mathematical models we use to reproduce our observations in simulations. Ideally, in a perfectly homogeneous cloud the collective instability would start from quantum fluctuations in the reverse mode, thermal excitations of this mode being completely frozen out at room temperature. However, thermal fluctuations in the atomic density distribution and, even more
important, spurious light scattering at the surfaces of the cavity mirrors scatter a certain amount of light into the reverse mode, which is sufficient to seed the instability. It is thus important to incorporate mirror backscattering in realistic theoretical models, as we will show in section~\ref{SecTheoryMirrors}. Section~\ref{SecExperiment} is devoted to presenting our experimental apparatus, the temporal sequence of an experimental run and several measurements. In particular, we will show the measured dependences of the collectively scattered light power on various parameters, such as atom number, pump power, and mirror backscattering. We will demonstrate that both regimes, the good- as well as the bad-cavity regime, can be realized and exhibit characteristic signatures. In section~\ref{SecExperimentTof}, we present and discuss time-of-flight absorption images taken on thermal and Bose-condensed atomic clouds. We conclude this paper with a discussion and a brief outlook.

\section{Theoretical background}
\label{SecTheory}

CARL and SRyS have been observed under very different experimental
circumstances and in different parameter regimes. In the case of
CARL the atoms are stored in a ring cavity, for SRyS they are held
in free space. CARL can be observed with $100~\mu$K cold atoms
\cite{Kruse03b}, while SRyS requires temperatures lower than
$1~\mu$K and is hardly seen with thermal clouds. Finally, CARL is
seen with pump laser detunings, which are 3 or 4 orders of magnitude
larger, than for SRyS.

Nevertheless, both phenomena have an important feature in common.
They share the same gain mechanism based on collective light
scattering and leading to an exponential instability in the atomic
density distribution and to the emission of coherent light pulses.
In this section, we will summarize and combine the main theoretical
results published in \cite{Inouye99, Piovella01b, Robb05} in order
to clarify the connection between CARL and SRyS in a consistent
picture supporting the understanding of our measurements. Later we
derive equations of motion valid in both regimes of CARL and SRyS.

\subsection{Self-amplification in CARL and SRyS}
\label{SecTheoryCarlSrys}

In the CARL experiments \cite{Kruse03b,Cube04,Slama07}, a cold or ultracold atomic cloud is brought into the mode volume of a unidirectionally pumped ring cavity. The pump light is very far detuned by more than $1~$nm. It is irrelevant whether the cloud is condensed or thermal. The atoms scatter light from the pumped into the reverse mode. Tiny fluctuations in the nearly homogeneous atomic density distribution are exponentially amplified. The atoms self-organize into a one-dimensional optical lattice and a red-detuned coherent probe light is emitted by the reverse mode.

\bigskip

The Rabi frequency generated by a single photon in the ring cavity of round-trip length $L$ and waist $w_0$ is $\Omega_1=\sqrt{3\Gamma c/k^2w_0^2L}$ \cite{Gangl00,Cube06}. The single-photon light-shift far from resonance, $U_0=\Omega_1^2/\Delta$, can also be interpreted as the Rabi frequency for the coupling between the pump and the probe mode, i.e.~the rate at which photons are exchanged between the modes.
The small signal gain can be derived from a linearization of the CARL equations \cite{Piovella01b,Robb05},
\begin{equation}\label{Eq01}
    G_c = \frac{2g^2N}{\kappa_{\mathrm{c}}}~,
\end{equation}
where $N$ is the atom number and
$\kappa_{\mathrm{c}}=\pi\delta_{\text{fsr}}/F$ the decay rate of the
light field in the cavity. $\delta_{\text{fsr}}$ is the free
spectral range of the cavity and $F$ its finesse. The quantity $g$
is given by
\begin{equation}\label{Eq02}
    g = \frac{\Omega_+\Omega_-}{2\Delta}~,
\end{equation}
where the Rabi frequency generated by the pump mode scales with the root of the pump photon numbers, $\Omega_+=\Omega_1\sqrt{n_+}$. The coupling strength in the probe mode is $\Omega_-=\Omega_1$.
>From the above equations, we get
\begin{equation}\label{Eq03}
    G_c = \frac{\Omega_+^2}{2\Delta}\frac{N}{\kappa_{\mathrm{c}}}\frac{\Omega_-^2}{\Delta}~.
\end{equation}

\bigskip

In the SRyS experiments performed up to date \cite{Inouye99,Kozuma99,Schneble03,Fallani05,Yoshikawa05}, an ultracold, in general Bose-condensed atomic cloud with ellipsoidal shape is irradiated by a short pump laser pulse modestly detuned from an atomic resonance by about $1~$GHz. The pulsed pump light drives a transient dynamics simultaneously forming a matter wave grating and emitting an optical mode into the BEC's long axis, which exponentially amplify each other.

Following Ref.~\cite{Inouye99}, one may associate the part of the BEC that corresponds to atoms which have scattered a photon with an atom number $N_r$. The remaining part consists of $N$ atoms. The density is modulated by interference between the two parts of the wave function, and the number of atoms that form the density modulation is $N_{\text{mod}}\propto\sqrt{2N N_r}$. As for usual Bragg scattering or Dicke superradiance the number of photons $n$ scattered at the density modulation is $n\propto N_{\text{mod}}^2\propto N_r$. Since every scattered photon generates a recoiling atom, the number of recoiling atoms increases like $\dot{N_r}\propto n$, and we get $\dot{N_r}=G_{\text{sr}}N_r$, i.e.~an exponential increase of recoiling atoms with a gain factor $G_{\text{sr}}$. This increase is mirrored by an identical rise of the number of scattered photons, which results in a gain mechanism for the scattered light mode. The incident and the scattered light mode are coherently coupled, just like in the case of CARL, so that in principle the scattered photons can be scattered back into the incident mode.

The superradiant gain can be expressed as
\begin{equation}\label{Eq04}
    G_{\text{sr}} = RN_0\frac{\Phi_s}{8\pi/3}~,
\end{equation}
where $R=\Gamma\Omega_+^2/(4\Delta^2+2\Omega_+^2+\Gamma^2)$ is the single-atom Rayleigh scattering rate, with $\Gamma$ being the linewidth of the atomic resonance, $\Delta$ the detuning, and $\Omega_+$ the Rabi frequency generated by the incident laser beam. $\Phi_s\simeq\lambda^2/\frac{\pi}{4}w^2$ is the scattering solid angle, with $w$ being the waist of the condensate. Hence, far from resonance,
\begin{equation}\label{Eq05}
    G_{\text{sr}} = \frac{\Omega_+^2}{\Delta^2}N_0\frac{3\Gamma}{2k^2w^2}~.
\end{equation}
This result can be brought into the same form as the CARL
gain~(\ref{Eq03}), if we interpret the condensate, whose length
along the long axis is $L$, as a cavity with free spectral range
$\delta_{\text{fsr}}=c/L$ and finesse $F_{\text{sr}}=\pi$. With this
interpretation the decay rate of the light mode scattered by the
condensate is given by the residence time of the light within the
BEC \cite{Stamper-Kurn00},
$\kappa_{\text{fsr}}=\pi\delta_{\text{fsr}}/F_{\text{sr}}=c/L$.
\begin{equation}\label{Eq06}
    G_{\text{sr}} = \frac{\Omega_+^2}{2\Delta^2}\frac{N_0}{\kappa_{\text{sr}}}\frac{3\Gamma\delta_{\text{fsr}}}{k^2w^2}
        = \frac{\Omega_+^2}{2\Delta}\frac{N_0}{\kappa_{\text{sr}}}\frac{\Omega_1^2}{\Delta}~.
\end{equation}
This result shows the equivalence of the superradiant gain and the gain occurring in CARL in equation (\ref{Eq03}).

\bigskip

The formal identity of the small signal gain of CARL and SRyS points
to the same roots of both phenomena. Nevertheless, their respective
experimental circumstances are quite different. The differences
become most apparent in the simultaneous build-up of the atomic
density grating and optical standing wave, occurring as well in CARL
as in SRyS. The difference lies in the storage of the coherence,
which is crucial in order to sustain the build-up process. In
principle the coherence can either be stored as a matter wave
coherence between different atomic momentum states or as a phase
coherence between the two involved light fields. In SRyS the optical
coherence time alone would be very small, as can be estimated from
the decay rates of the optical modes, which are on the order of
$\kappa_{\text{sr}}\simeq10^{12}~\text{s}^{-1}$. The coherence must
therefore be maintained in the atomic momentum states which then
form a matter wave grating. This is the reason why SRyS is very
sensitive to the temperature of the atomic cloud. The thermal energy
of the atoms must be smaller than the recoil energy
$k_BT<\hbar\omega_{\textrm{r}}=2\hbar^2k^2/m$. Otherwise, the
Doppler broadening leads to decoherence of the momentum states and
detroys the matter wave coherence and the  resulting density
grating.

For CARL the situation is reversed. CARL has been observed with
temperatures much higher than the recoil temperature, i.e.~in a
regime where interferences between atoms in Raman superpositions of
momentum states are quickly smeared out by Doppler broadening. Here,
the optical cavity plays the crucial role, because it
phase-coherently stores the participating light fields for times on
the order of several $\mu\text{s}$, given by the cavity decay rate
$\kappa_{\mathrm{c}}/2\pi=20~\text{kHz}\ll\kappa_{\text{sr}}$ which
is 7 orders of magnitude smaller than in the case of SRyS without
cavity.

\subsection{Collective gain in various regimes}
\label{SecTheoryRegimes}

The important point is now, that the broad range in which the collective gain can be varied in our experiment allows us to study CARL and SRyS dynamics as two opposite regimes of one system, called the good-cavity and the bad-cavity regime. Both regimes can be further divided into a semiclassical and a quantum domain and are characterized by two parameters, the CARL parameter $\rho$ and the scaled decay rate $\kappa$ \cite{Piovella01}. The CARL parameter is given by the product of the small signal gain and the decay rate of light both in units of the recoil frequency $\omega_{\textrm{r}}=2\hbar k^2/m$
\begin{equation}\label{Eqrho}
    \rho^3=\frac{G_{\textrm{c,sr}}}{\omega_{\textrm{r}}}\cdot\frac{\kappa_{\textrm{c,sr}}}{\omega_{\textrm{r}}}~.
\end{equation}
The scaled decay rate $\kappa=\kappa_{\textrm{c,sr}}/\omega_{\textrm{r}}\rho$ depends via $\rho$ on the gain, too. The good-cavity regime is given by $\kappa<1$, the bad-cavity regime by $\kappa>1$.\\

\begin{figure}[ht]
        \centerline{\scalebox{0.65}{\includegraphics{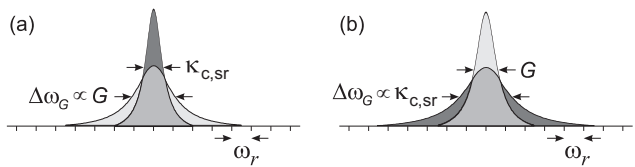}}}\caption{
        Representation of the two limiting cases of a long and short cavity lifetime. Shown are the cavity transmission profile
        (dark shaded areas) and the gain profile (bright shaded areas). (a) When the cavity linewidth is smaller than the gain,
        the good-cavity limit is realized. (b) When the gain is smaller, the superradiant limit is realized.}
        \label{fig:Scheme}
    \end{figure}

For the interpretation it is helpful to link the gain $G_{\textrm{c,sr}}$ to the gain bandwidth $\Delta\omega_G$, which is defined as the width of spectral range where the light scattering is exponentially amplified \cite{Piovella01}.
Let us first consider the semiclassical regime. The good-cavity limit is reached for strong saturation of the transition between the coupled cavity modes. This means that the gain, which can be interpreted as Rabi frequency, overwhelms the cavity decay width, $G_{\textrm{c,sr}}\gg\kappa_{\textrm{c,sr}}$. In this regime the transition is power-broadened by an amount $\Delta\omega_G\sim\omega_{\textrm{r}}\rho$ (see Fig. \ref{fig:Scheme}). This refers to the CARL experiments performed so far, where the gain bandwidth is proportional to the CARL parameter. In contrast, the bad-cavity regime is reached for small gain, $G_{\textrm{c,sr}}\ll\kappa_{\textrm{c,sr}}$. In this case, the gain bandwidth is given by the cavity decay rate, $\Delta\omega_G\sim\kappa_{\textrm{c,sr}}$. Obviously, the resolution of the gain profile cannot be better than $\kappa_{\textrm{c,sr}}$. This is the typical situation of SRyS.

\bigskip

The distinction between semiclassical and quantum regime is based on
the characteristic scale set by the recoil frequency
$\omega_{\textrm{r}}$. In the semiclassical regime the gain
bandwidth is large enough to amplify many adjacent momentum states
of the quantized motion $\Delta\omega_G>\omega_{\textrm{r}}$,
whereas in the quantum regime only one momentum state can be
amplified at a time $\Delta\omega_G<\omega_{\textrm{r}}$. Both, the
semiclassical as well as the quantum regime have been studied in
Ref.~\cite{Schneble03} in the bad-cavity limit by varying the gain
bandwidth. Strictly the CARL gain (\ref{Eq01})-(\ref{Eq03}) is only
valid in the quantum regime. In this regime the equivalence to the
SRyS gain (\ref{Eq04})-(\ref{Eq06}) appears in its clearest way. In
the semiclassical regime valid for our experiment the CARL gain is
reduced \cite{Robb05}, as has also been observed in SRyS
\cite{Schneble03}. In our experiment, the quantum limit could be
reached by reducing atom number and pump power. This would however
generate signals which are below the detection limit of our current
setup. Nevertheless, small deviations due to the quantum nature of
the atomic motion are expected, as will be briefly discussed in the
next sections.

\subsection{Equations of motion for atoms in a ring cavity}
\label{SecTheoryRegimes}

The system under consideration consists of ultra-cold or Bose-condensed atoms interacting with two counter-propagating modes of an optical cavity. The most general approach would treat all modes as quantized, in particular the atomic cloud would be described by a second-quantized matter wave field \cite{MooreMG99b,Piovella03}. Such an approach is necessary whenever mean field interactions or quantum statistical effects, like non-local interparticle correlations, particle fluctuations or entanglement, play a role. In the circumstances of our experiments, however, several simplifications can be made.

1.~All electronically excited states may be adiabatically eliminated \cite{Bonifacio95,Gangl00}. The detuning of the pump laser beam from the nearest resonance frequencies of the rubidium atom is so large, that the internal dynamics is continuously at a steady state keeping the population of the excited states at a negligible level. 2.~Propagation effects of light inside the atomic cloud \cite{Bonifacio97b,Zobay06} do not need to be considered. In comparison with the SRyS experiments, where the pump light is generally detuned by amounts on the order of $1~$GHz, our experiment uses $1000$ times larger detunings. Hence, the optical density of our atomic clouds at these detunings is negligibly small. 3.~Quantum statistical effects, such as entanglement, are predicted to occur naturally as a result of CARL dynamics \cite{Piovella03}. However, our experiment is not sensitive to signatures arising from quantum statistics. 4.~We treat all light fields classically. The mode volume of our cavity is of a size that the atom-field coupling constant larger than the cavity decay width, but it is much smaller than the spontaneous emission decay width of the atomic transition. Hence we are far from the cavity QED regime. Even in situations where shot noise could play a role, e.g.~in seeding the instability, perturbations arising from experimental imperfections (mirror backscattering) dominate. 5.~We treat the problem in one dimension, i.e.~along the optical axis of the cavity. Transversal oscillations of the atomic cloud, which may result from the collective dynamics \cite{Elsaesser03b} are not considered here. 6.~We neglect the backaction of the atoms on the pump light field (undepleted pump approximation). This is possible because the probe light is typically three orders of magnitude weaker than the pump field. In the experiment, the pump laser is tightly phase-locked to a cavity eigenfrequency. Consequently, as pointed out in Ref.~\cite{Kruse03b}, we can suppose a fixed phase relation between the incident pump laser field (labeled by the electric field amplitude normalized to the field generated by a single photon), $\alpha_{in}$, and the pumped cavity mode, $\alpha_+=\alpha_{in}\sqrt{\delta_{fsr}/\kappa_c}$. The phase can be arbitrarily chosen, e.g.~$\alpha_+$ can be taken as real.\\

Even though quantum statistical effects do not emerge from our measurements at temperatures close to or below the recoil limit the quantized nature of the atoms' motion influences their dynamics, as described by a model derived by Piovella and coworkers \cite{Piovella01}. Within this model and in the approximations specified above, the CARL Hamiltonian for an ensemble of $N$ atoms reads
\begin{align}\label{Eq11}
    H  & = \frac{1}{2m}\sum_{j=1}^N\hat{p}_j^2+\hbar\Delta_c\left(|\alpha_-|^2+|\alpha_+|^2\right)\\
        & + \hbar U_0\alpha_+\sum_{j=1}^N\left(\alpha_-^*e^{-2ik\hat{z}_j}+h.c.\right)~\notag,
\end{align}
where $U_0$ is the single-photon light shift, and $\Delta_c$ the detuning between pump and probe. The motional degrees of freedom, i.e.~the position $\hat{z}_j$ and the momentum $\hat{p}_j$ of every atom, satisfy the following commutation relation $[\hat{z}_j,\hat{p}_{j^{\prime}}] =i\hbar \delta_{jj^{\prime}}$. From the Heisenberg equations $i\hbar\dot{\hat{z}}=[\hat{z},H]$ and $i\hbar\dot{\hat{p}}=[\hat{p},H]$ we derive the equations of motion for the coupled system,
\begin{align}\label{Eq12}
    \frac{d\hat{z}_j}{dt} & =\frac{\hat{p}_j}{m}~,\\
    \frac{d\hat{p}_j}{dt} & =-2i\hbar kU_0\alpha_+\left(\alpha_-^*e^{2ik\hat{z}_j}-\alpha_-e^{-2ik\hat{z}_j}\right)~,\notag\\
    \frac{d\alpha_-}{dt} & =-(\kappa_c+i\Delta_c)\alpha_--iU_0\alpha_+\sum_{j=1}^Ne^{-2ik\hat{z}_j}~.\notag
\end{align}
In the last equation cavity damping has been introduced phenomenologically. $\hat{b}\equiv N^{-1}\sum\nolimits_je^{-2ik\hat{z}_j}$ measures the degree of atomic bunching. Starting from these equations, we either treat the motion classically or quantized \cite{Piovella01,Slama-07}. In the first case, we simply replace the position and momentum operators by their classical expectation values. These are the basic equations used to model most of the curves shown in this paper \cite{Perrin02}.

In order to check, whether quantum effects of the motion have an impact on the collective dynamics, we have derived from (\ref{Eq11}) a master equation for the density operator defining a momentum basis $\left|n\right\rangle_j$ such that $\hat{p}_j\left|n\right\rangle_j=2\hbar kn\left|n\right\rangle_j$ and $\left|\psi(\theta_j\right\rangle =\sum_nc_j(n)\left|n\right\rangle_j$. The calculations, which are analogous to those presented in Ref.~\cite{Piovella01}, are not reproduced here. They basically show that, for the parameters used in our experiments, quantum effects of the atomic motion are small. I.e., ~using the terminology of Ref.~\cite{Piovella01}, we are in the semiclassical regime.

\subsection{Modeling mirror backscattering and radiation pressure}
\label{SecTheoryMirrors}

Perturbative effects resulting from backscattering from the mirror
surfaces and from radiation pressure have been neglected so far.
Unfortunately, we found both effects to influence the experimental
observations, so that this idealization has to be given up. Let us
first discuss mirror backscattering. Dust particles or
irregularities on the mirror surfaces can scatter light from a
cavity mode into the counterpropagating mode. This effect is
well-known in laser gyroscopes, where it leads to phase-locking.
Interestingly, the effect is the more pronounced the better the
reflectivity of the mirrors and hence the finesse of the cavity
\cite{Cube06}. In principle, to describe mirror backscattering, one
has to know the precise locations of the scatterers on the mirrors.
As we explain in another paper \cite{Krenz07}, we can describe their
influence by a single scatterer localized at position
$z_{\textrm{s}}$ with a wavelength-dependent scattering rate
$U_{\textrm{s}}$. The scattering can be modeled in the very same way
as backscattering from atoms, except for the fact that the
scatterers are now fixed in space. Hence, we may just replace the
Hamiltonian~(\ref{Eq11}) by
\begin{equation}\label{Eq15}
    H' = H+\hbar U_{\textrm{s}}\alpha_+\left(\alpha_-^*e^{-2ikz_{\textrm{s}}}+h.c.\right)~.
\end{equation}
The resulting modified equations of motion are only changed by an additional term for the evolution of the field amplitude. I.e.~the third of the equations~(\ref{Eq12}) is supplemented with a gain rate $iU_{\textrm{s}}\alpha_+$ for the probe mode resulting from photons scattered out of the pump mode by mirror backscattering. In the experiment, we determine the amount of mirror backscattering $U_{\textrm{s}}$ from independent measurements.

\bigskip

Radiation pressure is due to spurious population of electronically
excited states under the influence of the pump laser beam. Although,
far from resonance the effect is weak, it still leads to a
noticeable acceleration of the atoms. Gangl and Ritsch
\cite{Gangl00} have shown that the adiabatic elimination of
electronically excited states introduces additional contributions in
the classical CARL equations scaling with the Rayleigh scattering
rate $\gamma_0$. This describes the effect of recoil heating due to
radiation pressure
\begin{align}\label{Eq16}
    m\frac{d^2z_j}{dt^2} & = -\hbar k\gamma_0\left(|\alpha_+|^2-|\alpha_-|^2\right)\\
    & - 2i\hbar kU_0\alpha_+\left(\alpha_-e^{2ikz_j}-\alpha_-^*e^{-2ikz_j}\right)~,\notag\\
    \frac{d\alpha_-}{dt} & = -(\kappa_c+N\gamma_0)\alpha_--N(\gamma_0+iU_0)\alpha_+b-iU_{\textrm{s}}\alpha_+~.\notag
\end{align}
The additional contributions not only lead to losses for the light
mode, but also exert an accelerating force onto the atoms.
Experimentally, we observe a broadening of the momentum distribution
by recoil heating which slightly impairs the collective dynamics for
measuring times longer than $100~\mu$s.

\section{Measurements}
\label{SecExperiment}
    \begin{figure}[ht]
        \centerline{\scalebox{0.55}{\includegraphics{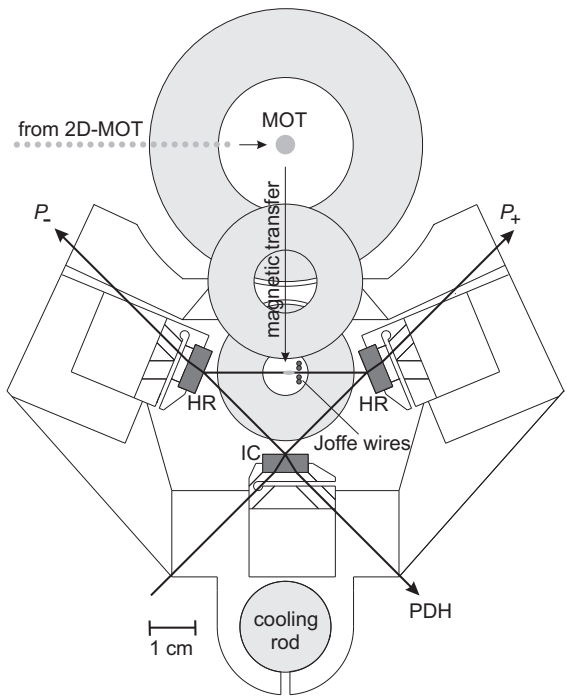}}}
        \caption{Technical drawing of the setup in the main chamber including coils for magnetic and magneto-optical trapping,
            wires for a Joffe-Pritchard type trap and the ring cavity. All pieces are held together by massive copper parts
            omitted in this figure for clarity.}
        \label{fig:setup}
        \end{figure}
We describe our experimental setup tracking the temporal sequence of an experimental run. The whole setup (shown in Fig.~\ref{fig:setup}) consisting of magnetic coils, wires and the ring cavity is placed inside a ultra-high vacuum chamber pumped by a cryogenic titanium sublimation pump and a $20$\,l/s ion getter pump to a pressure of about $10^{-11}$\,mbar. Heat produced in coils and wires inside the vacuum is dissipated via a temperature-stabilized cooling rod to a liquid nitrogen reservoir. A second vacuum chamber is connected with this main chamber via a differential pumping hole and contains a Rb partial pressure of several $10^{-7}$\,mbar. The second chamber accommodates a two-dimensional magneto-optical trap (2D-MOT) producing a cold atomic beam directed into the main chamber. From this atomic beam about $10^8$ atoms/s are recaptured in a standard magneto-optical trap (MOT) in the main chamber. After the MOT has been loaded for 15~s, the atoms are transferred into a magnetic trap produced by the same coils as the MOT. On a typical day, we load about $2\times10^8$ atoms at a temperature of $T=100~\mu\textrm{K}$ into the magnetic trap. The atoms are then magnetically transferred via a second into a third pair of coils, whereby the atoms are compressed adiabatically. The magnetic quadrupole field gradient between the third pair of coils is $160$\,G/cm in the horizontal and $320$\,G/cm in the vertical direction. With two pairs of wires separated by $1$\,mm and running parallel to the symmetry axis of the coils a Joffe-Pritchard type potential is created \cite{Silber05}. Typical values of the oscillation frequencies in this trap are $\omega_{\mathrm{r}}/2\pi=200$\,Hz and $\omega_z/2\pi=50$\,Hz at a magnetic offset field of $B_0=2$\,G with the $z$-direction pointing along the cavity mode through the gap between the wires. The vertical position of the wire trap can easily be shifted by the currents in the quadrupole coils. Inside the wire trap the atoms are cooled by forced evaporation: a microwave frequency is tuned resonantly to the ground state hyperfine structure and couples the trapped Zeeman state $|2,2\rangle$ and the untrapped state $|1,1\rangle$. We ramp down the frequency for 15~s starting from a detuning of $210$\,MHz and reach quantum degeneracy at a detuning of about $4$\,MHz with about $N=5\times10^5$ atoms at $T_c=800$\,nK. Almost pure condensates of $N=2\times10^5$ atoms can be achieved by ramping down to even lower frequencies. When the evaporative cooling stage is completed, the cold atoms are vertically transferred into the mode volume of the ring cavity. The ring cavity consists of one plane (IC) and two curved (HR) mirrors with a curvature radius of $R_c=10$\,cm. The round-trip length of the cavity is $8.5$\,cm, corresponding to a free spectral range of $\delta_{\textrm{fsr}}=3.5$\,GHz. One of the two counterpropagating modes is continuously pumped by a titanium-sapphire laser. The laser can be stabilized to this mode using the Pound-Drever-Hall (PDH) method. The quality factor of the cavity depends on the polarization of the incoupled light. For p-polarized light, a finesse of $F=87000$ is determined from a measured intensity decay time of $\tau=3.8~\mu\textrm{s}$. For s-polarized light the finesse is 6400.

\subsection{Experimental procedure}
\label{SecExperimentProcedure}

The measurements are performed in the following way. A cloud of cold atoms is magnetically transferred into the cavity. During this time the cavity is not pumped with light in order to prevent losses of atoms due to Rayleigh scattering.
    \begin{figure}[ht]
        \centerline{\scalebox{0.55}{\includegraphics{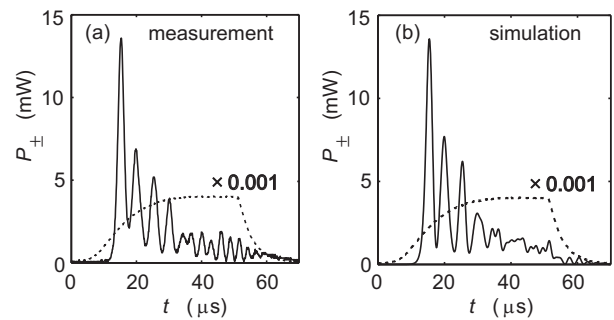}}}
        \caption{(a) Typical measured time signal (solid line) of the probe-light power. Experimental parameters are $N=1.5\cdot10^6$, $P_+=4$\,W,
            $\lambda=797.3$\,nm and $F=87000$. For visibility the pump-light power (dashed line) is scaled down by a factor of $0.001$.
            (b) Simulation of the CARL equations. The measured rise of the pump-light power is used in the simulation and
            the experimental parameters are fitted in order to agree with the measured time curve. The fitted
            parameters ($P_+$, $N$) are in reasonable agreement with
            the measured parameters.}
        \label{fig:timesignal}
    \end{figure}
This implies that the frequency of the laser cannot be stabilized to
a mode of the cavity during the transfer. As soon as the atoms are
inside the cavity, we switch on the pump light again and ramp its
frequency across the cavity resonance. This is done by means of a
piezo-electric transducer normally controlled by the slow branch of
the Pound-Drever-Hall (PDH) servo, which is interrupted for this
reason. As soon as the frequency is close to the cavity resonance,
the fast branch of the PDH servo acting on an acousto-optic
modulator (AOM) quickly pulls the laser frequency to the center of
the resonance and tightly locks its phase, thus compensating for the
frequency ramp. After a time of about $50~\mu$s the pump light is
turned off. The build-up time for the ring cavity pump mode is
limited by the bandwidth of the locking servo to about
$\tau_{bw}=20~\mu$s, which is longer than the cavity decay time.

As soon as the pump mode power builds up in the ring cavity, the
collective dynamics results in light scattering into the cavity
probe mode. The limited build-up time of the pump power leads to a
delayed and slightly weaker dynamics as compared to a rapid
switch-on.We study this dynamics mainly via the evolution of the
recorded probe light power $P_-$. The time signal of the probe light
shows characteristic maxima and minima like the ones presented in
Fig.~\ref{fig:timesignal}. This behavior can be explained most
easily in the case, where the atoms occupy a initial momentum
eigenstate and are coupled by the coherent dynamics to a final
momentum state. The temporal evolution is a Rabi oscillation-like
change of occupation from the initial to the final state. This
causes the build-up of an atomic density grating which reaches its
maximum with half of the atoms in each state and zero contrast when
all atoms are in the initial or the final state. The scattered light
is proportional to this density grating contrast. Maxima in the
probe light power therefore occur with each change of the momentum
state. In the situation depicted in Fig.~\ref{fig:timesignal} the
dynamics leads to the simultaneous occupation of an increasing
number of momentum states. The maximum atomic density grating washes
out with time and we observe a decrease of the light power maxima.

In the following we analyze the probe light power reached at the
first maximum $P_{-,1}$, because it shows a clear dependence on atom
number $N$, pump light power $P_+$, laser wavelength $\lambda$,
finesse of the cavity $F$, and on the atomic cloud's temperature
$T$. In contrast, it is quite robust against perturbative effects
such as mirror backscattering. Simulations of the CARL dynamics like
shown in Fig.~\ref{fig:timesignal}(b) are performed by numeric
integration of (\ref{Eq16}) with the explicit Euler method. We
simulate the trajectories of $N_s=100$ atoms, each representing
$N/N_s$ real atoms. At the beginning of the simulation the atoms are
spread in position over half a wavelength with equal spacings. For
simulations of clouds with temperature $T=0$ the start momentum of
all atoms is set to $p_j=0$. For simulations of clouds with nonzero
temperature the momenta at the beginning are normally distributed
with $\langle p_j^2\rangle=mk_BT$.

\subsection{Mirror backscattering}
\label{SecExperimentMirrors}

\begin{figure}[ht]
        \centerline{\scalebox{0.55}{\includegraphics{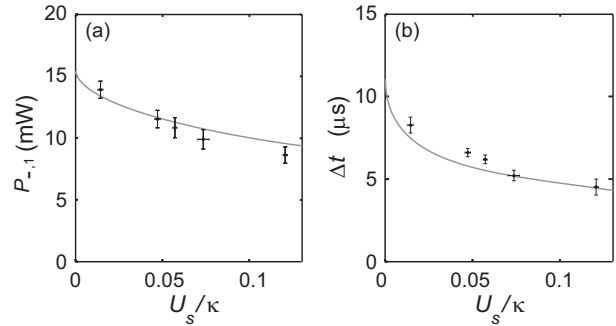}}}
        \caption{Influence of mirror backscattering (a) on the peak probe power $P_{-,1}$ and (b) on the delay of this peak $\Delta t$.
            The rate of mirror backscattering $U_{\textrm{s}}$ is given in units of the cavity decay rate $\kappa$. Measurements (circles) and
            simulations (lines) show a decrease of both $P_{-,1}$ and $\Delta t$ with rising $U_{\textrm{s}}$. The experimental parameters are
            $\lambda=796.1$\,nm, $N=1.5\times 10^6$, $P_+=500$\,mW and $F=87000$. In the simulations atom number and pump
            power are fitted to $1.9~N$ and $0.6~P_+$ in order to reach good quantitative agreement with the
            measurements. Good qualitative agreement is also reached if the simulations are performed with the measured
            parameters.}
        \label{fig:spr}
    \end{figure}

Scattering from the mirror surfaces leads to the presence of light
in the probe mode even in the absence of atoms in the cavity. In the
presence of atoms, this light influences the atomic collective
dynamics. Fig.~\ref{fig:spr} shows the impact of mirror
backscattering on the height of the first maximum $P_{-,1}$ and on
the time delay $\Delta t$ from switching on the pump until the
maximum is reached.

The backscattering rate strongly depends on the wavelength of the
pump laser, when it is resonant to an eigenfrequency of the cavity
\cite{Krenz07}. This phenomenon can be understood as interference of
the waves backscattered from all three cavity mirrors. From the
experimental point of view, the most interesting feature is that
backscattering can be avoided by a proper choice of the resonant
cavity mode. The mirror-induced probe light power varies between
almost 0 and 0.6\% of the pump power.

Backscattered light in the probe mode represents an artificial
instability, which seeds the collective dynamics. Consequently,
increased mirror backscattering reduces the time delay $\Delta t$.
On the other hand, the maximum probe light power $P_{-,1}$ decreases
with $\Delta t$, because the finite switch-on time limits the pump
light available at this stage. This behavior is verified by the
measurements shown in Fig.~\ref{fig:spr}. For these measurements we
vary the mirror backscattering by choosing different longitudinal
cavity modes \cite{Krenz07}.

In the simulations shown in same figure, the finite switch-on time
is taken into account. The atom number and pump power are fitted in
order to reach good agreement with the experimental data, but the
general behavior can be reproduced without free parameters. For a
hypothetic sudden switch-on, we would expect a much weaker
dependence of $P_{-,1}$ on mirror backscattering.

The observation that increased mirror backscattering leads to a
faster rise of the collective dynamics only applies, when the amount
of mirror backscattering is smaller than the atomic coupling
strength $U_{\textrm{s}}<NU_0$, which is true for the above given
values. For a reduced atom number of about $N\sim 10^5$ though,
mirror backscattering is on the same order of magnitude as the
atomic coupling. In this case, it is able to suppress the collective
dynamics, which we do observe experimentally. When we use
Bose-Einstein condensed clouds, atom numbers are precisely on the
order of $10^5$. It is therefore necessary to resort to cavity modes
with ultra-low mirror backscattering. To control and cancel the
amount of mirror backscattering, we have developed a method
described in \cite{Krenz07} based on the injection of an additional
light field into the probe mode of the cavity.

\subsection{Pump power}
\label{SecExperimentPump}

The dynamics of the collective instability depends on the pump light power. A reduction of pump power leads to a decrease of the contrast of the optical standing wave resulting from the interference of the pump and probe modes.
\begin{figure}[ht]
        \centerline{\scalebox{0.55}{\includegraphics{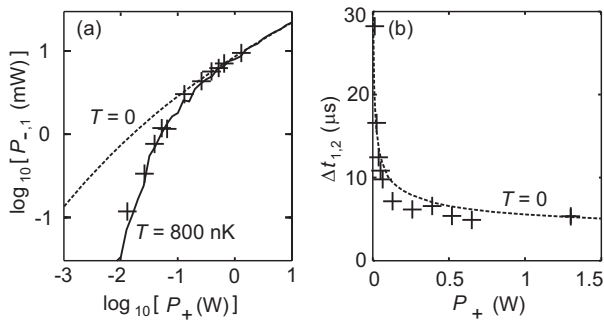}}}
        \caption{Influence of the pump light power on (a) the peak probe power $P_{-,1}$ and (b) the time delay between the first and the
            second superradiant peak $\Delta t_{1,2}$. Simulations are without free parameters. Residual fluctuations in the simulation for
            $T=800$\,nK in (a) are due to a limited simulated atom number. Experimental parameters are $\lambda=796.1$\,nm, $N=2.4\times10^6$,
            and $F=87000$. The stochastic error lies within the size of the markers.}
        \label{fig:pdep}
    \end{figure}
This weakens the collective dynamics. In previous experiments \cite{Cube04}, where the CARL has been exposed to the dissipative and diffusive forces of an optical molasses, we observed a threshold behavior in the pump power. In contrast, the present setup lacks a strongly dissipative reservoir, so that it is unclear whether CARL with BECs can show a threshold behavior. The only channel available to dissipation in this setup is transmission through the cavity mirrors. This provides a coupling of the cavity modes to the electromagnetic field of the surroundings, which to good approximation can be regarded as a zero-temperature reservoir of photons. One therefore would expect dissipation without diffusion.

We observed that temperature effects can lead to a threshold-like behavior, if the atoms are not Bose-condensed. Fig.~\ref{fig:pdep}(a) shows measurements of the maximum probe light power $P_{-,1}$ as a function of the pump power $P_+$. The data agree very well with simulations (solid line) using the parameters specified in the captions of Fig.~\ref{fig:pdep} and a temperature of the atoms of $T=800$\,nK. The dotted line is a simulation with the same parameters, but at temperature $T=0$. Down to a pump power of about $P_+\approx0.1$\,W, both curves coincide. Below this value the probe power is considerably reduced if the temperature of the atoms is finite. This demonstrates that thermal motion of the atoms can suppress the collective dynamics if the gain is not strong enough \cite{Note01}.
    \begin{figure}[ht]
        \centerline{\scalebox{0.55}{\includegraphics{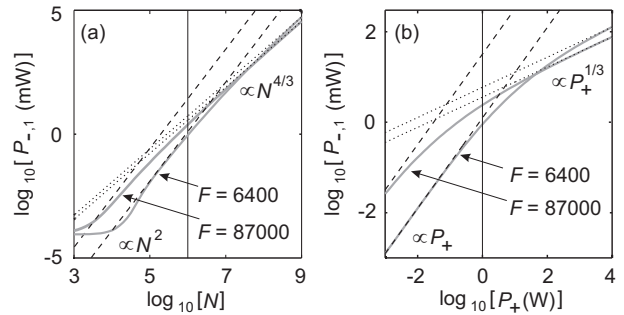}}}
        \caption{Simulations (solid lines) of the transition from bad-cavity to good-cavity regime with respect to (a) atom number and
            (b) pump power. Each dependency is plotted for two values of the finesse $F$. The experimental parameters are $P_+=1$\,W in
            (a) and $N=10^6$ in (b). For the $F=87000$ simulations the wavelength is $\lambda=796.1$\,nm for $F=6400$ it is
            $\lambda=795.3$\,nm. The vertical lines in each figure show the value of the parameter held fixed in the other part of the
            figure. They characterize the region where our experiments take place. The dotted (dashed) lines show the asymptotic behavior
            typical for the good-cavity (bad-cavity) regime. The deviation of the simulations from the asymptotic behavior [solid curves in
            Fig.~(a) below $N=10^4$] stems from mirror backscattering, which plays a major role for small atom numbers and suppresses the
            collective dynamics. The simulations are performed at $T=0$ in
            order to show the underlying physics without being
            influenced by temperature effects.}
        \label{fig:fdep}
    \end{figure}
Another observable which depends on the pump power is the time
difference $\Delta t_{1,2}$ between the first and the second
superradiant light pulse. This time difference corresponds to the
typical time-scale, on which the atomic momentum distribution is
shuffled between different momentum states. The stronger the pump
power is, the faster the momentum distribution changes. This
connection is shown in Fig.~\ref{fig:pdep} (b), where the data agree
very well with a simulation with the above given parameters and an
atomic temperature of $T=0$. A simulation with the realistic atomic
temperature of $T=800$\,nK hardly differs from the $T=0$ curve and
is omitted in Fig.~\ref{fig:pdep} (b) for clarity. This shows that
the time difference $\Delta t_{1,2}$ is quite insensitive to the
momentum spread of the atoms.

\subsection{Finesse}
\label{SecExperimentFinesse}

The CARL model comprises different regimes, which are denoted as good-cavity and bad-cavity regime. While former work in our group was performed in the good-cavity regime \cite{Kruse03,Cube04}, the SRyS experiments are very far in the bad-cavity regime \cite{Inouye99,Schneble03}. With our new apparatus we are able to reach both regimes by varying the finesse of the cavity and to find characteristic signatures of the regimes in the comportment on certain experimental parameters. The maximum probe light power scales in the good-cavity regime with $P_{-,1}\propto N^{4/3}\cdot P_+^{1/3}$ and in the bad-cavity regime with $P_{-,1}\propto N^2\cdot P_+$ \cite{Bonifacio94}. Which regime is reached does not only depend on the finesse $F$, but also on the atom number and the pump power themselves.
    \begin{figure}[ht]
        \centerline{\scalebox{0.55}{\includegraphics{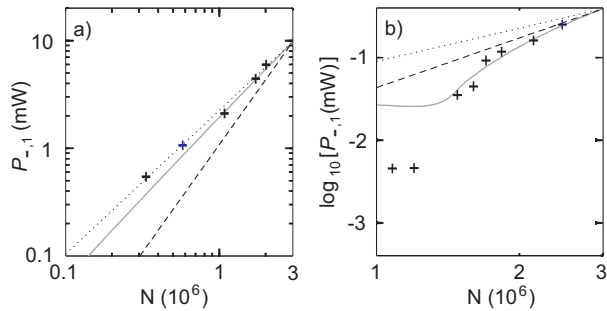}}}
        \caption{Measured dependency of the maximum probe light power as a function of atom number for values of the finesse of
            (a) $F=87000$ and $F=6400$. The parameters are a) $\lambda=796.1$\,nm, $P_+=1.43$\,W and b) $\lambda=795.3$\,nm, $P_+=66$\,mW.
            Comparing the data points to the asymptotic behavior shown in the dotted and dashed lines the situation in a) can be
            identified as good-cavity regime and the situation in (b) as bad-cavity regime. The behavior is confirmed by simulations
            with no free parameters (solid lines).
            The values of the data points are scaled by (a) 0.75 and (b) 2.8 in order to
            improve agreement with the simulation. This systematic error of the data points is due to uncertainties in the
            calibration of the probe light power. These depends on the polarisation of the light and for this reason
            we have to apply different scalings for low and good finesse. Nevertheless the dependency on atom number,
            which in the logarithmic plot shows up as a different slope is not changed by this pure
            multiplication. The stochastic error lies within the size of the markers.}
        \label{fig:ndep}
    \end{figure}
As discussed in Sec.~\ref{SecTheoryRegimes}, the regime is determined by the relative size of the cavity decay rate, $\kappa_{\mathrm{c}}\sim F^{-1}$, and the gain bandwidth which depends on the collective gain $G\sim nNU_0^2/\kappa_{\mathrm{c}}$. Hence, the good-cavity regime is characterized by large atom numbers and large pump powers, and the bad-cavity regime by small atom numbers and small pump powers. This feature is shown in Fig. \ref{fig:fdep}, where the dependence is simulated for the two values of the finesse accessible to our experiment. As can be seen, the transition between the two regimes is not sudden, but spreads across a wide range of atom number and pump power.

Measurements of the dependence of the maximum probe light power on atom number are shown in Fig.~\ref{fig:ndep}. The finesse of the ring cavity can be set to either $F=87000$ in Fig.~(a) or $F=6400$ in Fig.~(b) by simply rotating the polarization of the pump light with respect to the symmetry plane of the cavity. This enables us o probe both, the good-cavity and the bad-cavity regime. The asymptotic dependency in the good-cavity regime is shown by dotted lines, the dependency in the bad-cavity regime by dashed lines. The solid line represents a simulation with no free parameters. By varying the atom number in (a) between $N_1=3\cdot10^5$ and $N_2=2\cdot10^6$ the corresponding CARL parameters [Eq.~(\ref{Eqrho})] \cite{Bonifacio94} are $\rho_1=4.7$ and $\rho_2=7.0$, the corresponding scaled decay rates are $\kappa_1=\kappa_c/\omega_{\mathrm{r}}\rho_1=0.3$ and $\kappa_2=0.2$. The conditions $\kappa_{1,2}<1$ and $\rho_{1,2}>1$ are typical for the semi-classical good-cavity regime. Indeed, the data points are lying close to the good-cavity theoretical lines. In Fig.~\ref{fig:ndep}(b) the measured atom numbers between $N_3=1.1\cdot10^6$ and $N_4=2.5\cdot10^6$ correspond to CARL parameters between $\rho_3=5.1$ and $\rho_4=6.7$ and scaled decay rates between $\kappa_3=3.7$ and $\kappa_4=2.8$. The conditions $\kappa_{3,4}>1$ and $\rho_{3,4}>\kappa_{3,4}$ are typical for the semi-classical bad-cavity regime. This is confirmed by the data points which seem to be approximated by the good-cavity asymptotic line for high atom numbers. The discrepancy for low atom numbers is due to mirror backscattering. This effect is also visible in the simulation.

\subsection{Temperature}

With our apparatus the atomic temperature can be varied within a range from below one $\mu\textrm{K}$ to several tens of $\mu\textrm{K}$.
    \begin{figure}[ht]
        \centerline{\scalebox{0.55}{\includegraphics{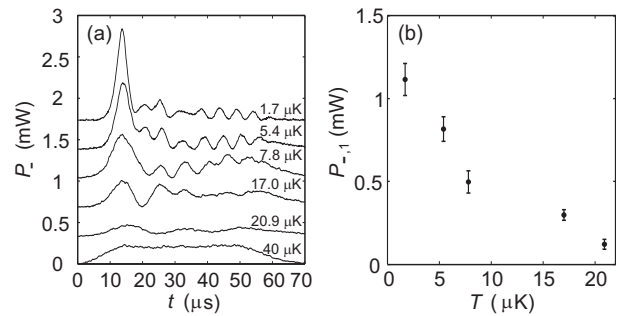}}}
        \caption{(a) Measured time signal of the probe-light for different atomic temperatures. For clarity the curves are shifted by
            $0.35$\,mW from each other. The experimental parameters are for all curves $N=10^6$, $\lambda=796.1$\,nm and $F=87000$.
            The signal decreases and the contrast is washed out for rising temperature.
            (b) Maximum probe-light power as a function of temperature extracted from (a).}
        \label{fig:tdep}
    \end{figure}
This allows us to systematically examine the influence of the temperature on the collective dynamics and identify the role of quantum statistics in the dynamics of CARL and SRyS. Fig.~\ref{fig:tdep}(a) presents recorded time signals of the probe light for different temperatures.

The curves show characteristic trains of superradiant pulses. With rising temperature the maximum probe-light power decreases and subsequent pulses are washed out. The bottom curve, which corresponds to a temperature of $T=40~\mu\textrm{K}$, shows no modulation of the light power and resembles the time evolution of pure mirror backscattering. The decrease of the maximum probe light power is separately plotted in (b). Obviously, a rising temperature leads to a suppression of the collective dynamics. This can be explained by the fact that the self-amplified optical standing wave has to arrange the atoms into an atomic grating. This is only possible if the depth of the optical lattice is larger than the thermal energy of the atoms. For that reason a rising temperature leads to fewer atoms participating in the gain mechanism. This is the reason why we cannot see CARL activity in the present experiment with atom numbers of $N=10^6$ at a temperature of $T=40~\mu\textrm{K}$, while we observed CARL in recent experiments with atom numbers of $N=10^7$ at temperatures well above $T=100~\mu\textrm{K}$ \cite{Kruse03b,Cube04}. The fact that CARL is observable at all with thermal clouds of atoms, is the proof that quantum statistical phenomena do not play a role for the dynamics of CARL.

\subsection{Evaluation of absorption images}
\label{SecExperimentTof}
    \begin{figure}[ht]
        \centerline{\scalebox{0.55}{\includegraphics{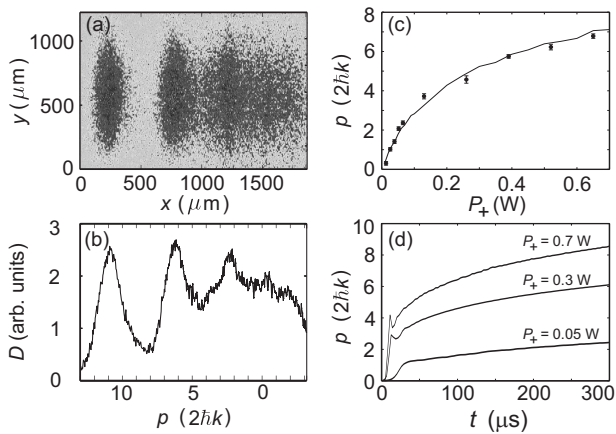}}}
        \caption{(a) Typical absorption image of a thermal atomic cloud after the CARL dynamics and $10$\,ms ballistic expansion.
            (b) Vertically integrated optical density of the cloud. (c) Measured mean momentum as a function of pump power
            compared to a simulation with no free parameter.
            (d) Simulated time evolution of the mean momentum. The momentum is given in units of the recoil momentum
            $p_{\mathrm{r}}=2\hbar k$. The experimental parameters are $\lambda=796.1$\,nm, $N=2.4\times10^6$ and
            $F=87000$. The atomic temperature is  $T\simeq 1$\,$\mu$K.}
        \label{fig:carlpic}
        \end{figure}
After a time period where the atoms are exposed to collective
dynamics, the atoms are released from the magnetic trap. The atomic
cloud expands ballistically, and after a time-of-flight of typically
$t_{\textrm{TOF}}=10$\,ms an absorption image is recorded, revealing
the momentum distribution of the atoms in the trap.
Fig.~\ref{fig:carlpic}(a) shows a typical image of a thermal atomic
cloud with (b) the vertically integrated optical density. The
momentum can be calculated from the horizontal displacement of the
atoms. Individual momentum states cannot be resolved, because the
momentum distribution appears broadened by the thermal motion.
Nevertheless, interesting information like the mean momentum
$\langle p\rangle$ can be extracted from such images. Therefore, we
calculate the center-of-mass of the vertically integrated optical
density. This mean momentum can be examined as a function of the
experimental parameters. Fig.~\ref{fig:carlpic}(c) shows this
dependency of the pump power. The measurements are very well
reproduced by simulations of the CARL equations. The simulations in
Fig.~\ref{fig:carlpic}(d) show that the mean momentum increases
rapidly during the first $T=50~\mu\textrm{s}$ and then starts to
saturate. The saturation is due to the presence of the optical
cavity restricting the range of accessible momentum states. In the
simulations, we assume a realistic temperature of
$T=1.2~\mu\textrm{K}$. The strong spatial modulation of the atomic
density in Fig.~\ref{fig:carlpic}(a) depicts the momentum
distribution generated by the collective dynamics. This behavior is
qualitatively supported by simulations.

\begin{figure}[ht]
        \centerline{\scalebox{0.55}{\includegraphics{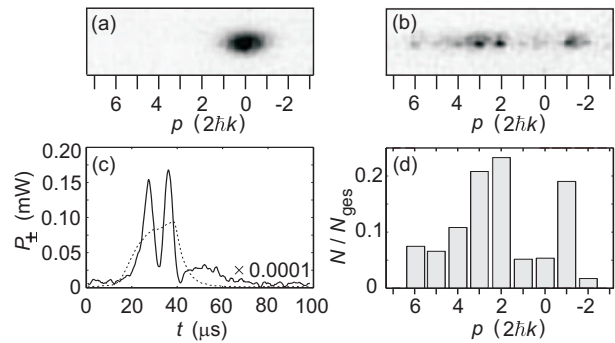}}}
        \caption{Absorption images of Bose-Einstein condensates after $10$\,ms ballistic expansion (a) without CARL and (b) with CARL
            activity. (c) Simultaneously recorded pump (dashed line) and probe power (solid line). Pump power is scaled down by $10^{-4}$.
            (d) Momentum distribution derived from Figure~(b).}
        \label{fig:beccarl}
    \end{figure}

If as shown in Fig.~\ref{fig:beccarl} a Bose-Einstein condensate is
used, we are able to resolve individual momentum states for (a) no
pump light-field and (b) a pump light-power of
$P_+^{\textrm{max}}\approx1$\,W. Due to the short interaction time
of the BEC with the light-field of $t_{\textrm{ia}}\approx
40~\mu\textrm{s}$ only two superradiant maxima are observed in (c).
The measured atomic momentum distribution after the interaction in
(d) shows a depopulation of the $|p\rangle=|0\rangle$ state and a
shift towards momentum states with positive momentum. The
substantial population of the momentum state with negative momentum
$|p\rangle=|-1\rangle$ is due to the semiclassical behavior of the
system and is equivalent to the observation of momentum spread in
\cite{Schneble03}.

\section{Conclusion}

We conclude this paper with the statement, that the collective
atomic recoil laser and superradiant Rayleigh scattering are two
faces of the same medal. Previous theoretical work
\cite{Bonifacio97} has shown that the characteristic quantity
distinguishing both effects is the collective gain bandwidth
compared to the cavity decay rate. Our experiment is designed to
give access to both regimes, the superradiant (or bad-cavity) regime
and the good-cavity regime. The observed characteristic dependence
of the instability amplitude on the atom number allows us to clearly
identify the regimes, and to experimentally demonstrate the
intrinsic link between both phenomena.

Another important result is the presence of collective instabilities
at high temperatures. In earlier experiments, CARL dynamics have
been observed with atomic clouds as hot as several $100~\mu$K
\cite{Kruse03b}. This proves that the gain process underlying both,
SRyS and CARL, is not based on quantum statistics, but on
cooperativity \cite{Moore01}. From this results a better
understanding of the intricate relationship between CARL and
superradiance.

This experiment represents the first study of Bose-Einstein
condensates in macroscopic cavities. For the experiments described
within this publication though, the quantum degeneracy of the atoms
is unimportant. However in future experiments, we want to study the
role of quantum statistics in a regime, where photonic and
matter-wave modes are coherently coupled \cite{Horak00}. In this new
regime the CARL dynamics may generate entangled states between atoms
and scattered photons \cite{MooreMG99,Piovella03}.

Another challenge would be to reach the so-called quantum limit.
This limit is distinguished from the semiclassical limit by the fact
that the gain bandwidth is so small,
$\Delta\omega_G\gg\omega_{\mathrm{r}}$, that only adjacent momentum
states of the atomic motion are coupled. This case (provided the
temperature is very low) results in a train of self-similar
superradiant pulses \cite{Piovella01b}. In our experiment this
regime could be reached by enhancing the finesse of the ring cavity
or by reducing $\omega_{\mathrm{r}}$, e.g.~by tuning the pump laser
to an atomic resonance at a much higher frequency. To treat this
regime the use of quantized atomic motion in the CARL equations is
compulsory \cite{Piovella01b}.

\bigskip

This work has been supported by the Deutsche Forschungsgemeinschaft
(DFG) under Contract No. Co~229/3-1. We like to thank W. Ketterle
for helpful discussions.


\begin{thebibliography}{10}

\bibitem{Dicke54}R.~H. Dicke, Phys. Rev. \textbf{93}, 99 (1954)

\bibitem{Skribanowitz73}N. Skribanowitz, I.~P. Hermann, J.~C. MacGilliwray, and M.~S. Feld, Phys. Rev. Lett. \textbf{30}, 309 (1973).

\bibitem{Moessbauer58}R. L. M\"o\ss bauer, Z. Physik \textbf{151}, 124 (1958).

\bibitem{Steane97}A. Steane, Appl. Phys. B \textbf{64}, 623 (1997).

\bibitem{Bonifacio94}R. Bonifacio and L.~De Salvo, Nucl. Instrum. Methods \textbf{341}, 360 (1994).

\bibitem{Bonifacio95}R. Bonifacio and L.~De Salvo, Appl. Phys. B \textbf{60}, S233 (1995).

\bibitem{Kruse03b}D. Kruse, Ch. von Cube, C. Zimmermann, and Ph.~W. Courteille, Phys. Rev. Lett. \textbf{91}, 183601 (2003).

\bibitem{Cube04}C. von Cube, {\em et al.}, Phys. Rev. Lett. \textbf{93}, 083601 (2004).

\bibitem{Slama07}S. Slama, S. Bux, G. Krenz, C. Zimmermann, and Ph.~W. Courteille, Phys. Rev. Lett. \textbf{98}, 053603 (2007).

\bibitem{Inouye99}S. Inouye, {\em et al.}, Science \textbf{285}, 571 (1999).

\bibitem{Kozuma99}M. Kozuma, {\em et al.}, Science \textbf{286}, 2309 (1999).

\bibitem{Fallani05}L. Fallani, {\em et al.}, Phys. Rev. A \textbf{71}, 033612 (2005).

\bibitem{Black03}A.~T. Black, H.~W. Chan, and V. Vuleti\'c, Phys. Rev. Lett. \textbf{91}, 203001 (2003).

\bibitem{Elsaesser03b}Th. Els\"asser, B. Nagorny, and A. Hemmerich, Phys. Rev. A \textbf{69}, 033403 (2003).

\bibitem{Labeyrie06}G. Labeyrie, F. Michaud, and R. Kaiser, Phys. Rev. Lett. \textbf{96}, 023003 (2006).

\bibitem{Moore01}M.~G. Moore and P. Meystre, Phys. Rev. Lett. \textbf{86}, 4199 (2001).

\bibitem{Ketterle01}W. Ketterle and S. Inouye, Phys. Rev. Lett. \textbf{86}, 4203 (2001).

\bibitem{Yoshikawa05}Y. Yoshikawa, Y. Torii and T. Kuga, Phys. Rev. Lett. \textbf{94}, 083602 (2005).

\bibitem{Piovella01}N. Piovella, R. Bonifacio, B.~W.~J. McNeil, and G.~R.~M. Robb, Opt. Commun. \textbf{187}, 165 (1997).

\bibitem{Piovella01b}N. Piovella, M. Gatelli, and R. Bonifacio, Opt. Commun. \textbf{194}, 167 (2001).

\bibitem{Robb05}G.~R.~M. Robb, N. Piovella, and R. Bonifacio, J. Opt. B: Quantum Semiclass. Opt. \textbf{7}, 93 (2005).

\bibitem{Gangl00}M. Gangl and H. Ritsch, Phys. Rev. A \textbf{61}, 043405 (2000).

\bibitem{Cube06}C. von Cube, {\em et al.}, Fortschr. Phys. \textbf{54}, 726 (2006).

\bibitem{Schneble03}D. Schneble, {\em et al.}, Science \textbf{300}, 475 (2003).

\bibitem{Stamper-Kurn00}D. M. Stamper-Kurn and W. Ketterle, Proc. Les Houches Summer School, Session LXXII.

\bibitem{MooreMG99b}M.~G. Moore, O. Zobay, and P. Meystre, Phys. Rev. A \textbf{60}, 1491 (1999).

\bibitem{Piovella03}N. Piovella, M. Cola, and R. Bonifacio, Phys. Rev. A \textbf{67}, 013817 (2003).

\bibitem{Bonifacio97b}R. Bonifacio, L. De Salvo, and G. R. M. Robb,  Opt. Comm. \textbf{137}, 276 (1997).

\bibitem{Zobay06}O. Zobay and G.~M. Nikolopoulos, Phys. Rev. A \textbf{73}, 013620 (2006).

\bibitem{Slama-07}S. Slama, Ph.D. thesis, Universit\"at T\"ubingen, 2007 (http://www.uni-tuebingen.de/ub/elib/tobias.htm).

\bibitem{Perrin02}M. Perrin, Zongxiong Ye, and L. M. Narducci, Phys. Rev. A \textbf{66}, 043809 (2002).

\bibitem{Krenz07}G. Krenz, S. Bux, S. Slama, and Ph.~C. Courteille, Appl. Phys. B \textbf{} in press (2007).

\bibitem{Silber05}C. Silber, {\em et al.}, Phys. Rev. Lett. \textbf{95}, 170408 (2005).

\bibitem{Note01}An interesting question in this context is, whether quantum fluctuations in a BEC could lead to stochastic and therefore diffusive processes, which then would cause a threshold behavior at zero temperature.

\bibitem{Kruse03}D. Kruse, {\em et al.}, Phys. Rev. A \textbf{67}, 051802(R) (2003).

\bibitem{Bonifacio97}R. Bonifacio, G.~R.~M. Robb, and B.~W.~J. McNeil, Phys. Rev. A \textbf{56}, 912 (1997).

\bibitem{Horak00}P. Horak, S.~M. Barnett, and H. Ritsch, Phys. Rev. A \textbf{61}, 033609 (2000).

\bibitem{MooreMG99}M.~G. Moore and P. Meystre, Phys. Rev. A \textbf{59}, R1754 (1999).

\end{thebibliography}
\end{document}